# Generation of COVID-19 Chest CT Scan Images using Generative Adversarial Networks

Prerak Mann[1]
*Computer Engineering*
*Delhi Technological University*
New Delhi, India
prerakmann_2k17co241@dtu.ac.in

Sahaj Jain[1]
*Computer Engineering*
*Delhi Technological University*
New Delhi, India
sahajjain_2k17co291@dtu.ac.in

Saurabh Mittal[1]
*Computer Engineering*
*Delhi Technological University*
New Delhi, India
saurabhmittal_2k17co309@dtu.ac.in

Aruna Bhat[2]
*Computer Science and Engineering*
*Delhi Technological University*
New Delhi, India
aruna.bhat@dtu.ac.in

*Abstract*—SARS-CoV-2, also known as COVID-19 or Coronavirus, is a viral contagious disease that is infected by a novel coronavirus, and has been rapidly spreading across the globe. It is very important to test and isolate people to reduce spread, and from here comes the need to do this quickly and efficiently. According to some studies, Chest-CT outperforms RT-PCR lab testing, which is the current standard, when diagnosing COVID-19 patients. Due to this, computer vision researchers have developed various deep learning systems that can predict COVID-19 using a Chest-CT scan correctly to a certain degree. The accuracy of these systems is limited since deep learning neural networks such as CNNs (Convolutional Neural Networks) need a significantly large quantity of data for training in order to produce good quality results. Since the disease is relatively recent and more focus has been on CXR (Chest XRay) images, the available chest CT Scan image dataset is much less. We propose a method, by utilizing GANs, to generate synthetic chest CT images of both positive and negative COVID-19 patients. Using a pre-built predictive model, we concluded that around 40% of the generated images are correctly predicted as COVID-19 positive. The dataset thus generated can be used to train a CNN-based classifier which can help determine COVID-19 in a patient with greater accuracy.

## I. INTRODUCTION

COVID-19 is on the spread, and without any known vaccine or treatment, a significant number (~10%) of people having fatal reactions, and a mortality rate of ~2-3%, there is an urgent need to test and isolate people. According to Chinese authorities' publications, the diagnosis of COVID-19 has to be verified by gene sequencing of respiratory or blood specimens or RT-PCR (reverse-transcription polymerase chain reaction). However, due to the limitations of transportation and sample collection, as well as the testing kit's performance, throat swab samples have RT-PCR's total positive rate to be approximately around 30% - 60% only. Another factor is that one has to wait for the lab test results, which usually takes around 24-36 hours.

In a study of more than 1000 patients[2], chest-CT outperformed RT-PCR lab testing when diagnosing COVID-19 patients. From the results, the chest CT scans of 88% of the patients were positive, while only 59% had positive RT-PCR results. This goes to show that chest CT scans are more accurate for the screening of the novel coronavirus disease.

The number of chest CT Scan images available for COVID-19 patients is very less due to which the accuracy of a Convolutional Neural Network (CovNet/CNN) classifier is limited. Deep learning neural networks such as CNNs (convolutional neural networks) need a significantly large number of data for training in order to produce good quality results. Since the disease is relatively recent and more focus has been on CXR (Chest XRay) images, the available chest CT Scan image dataset is much less.

In order to tackle this situation, we came up with a solution of increasing the available Chest-CT scan dataset using synthetic images generated by a GAN model. This extended dataset can now be used to develop an improved CNN-based classifier model.

## II. BACKGROUND AND RELATED WORK

Image generation is the task of generating new synthetic images from a given dataset. There are many machine learning techniques such as Variational Autoencoders (VAEs), Autoregressive model, Flow model, Hybrid Models (a combination of these techniques), Generative Adversarial Networks (GANs). The latest of these techniques is GANs [4]. They belong to a set of generative models and can be used to generate text, audio, and images. GANs have had their application increase manifold in the past few years in fields such as science, fashion, art, advertising, etc., and have an advantage over other techniques when the task is to generate images that are very realistic and virtually indistinguishable from real images.

GAN consists of two primary networks - Generator (G(z)) and Discriminator (D(x)). The generator module in GANs is used to create artificial samples of data by incorporating feedback from the discriminator. Its objective is to deceive the discriminator into classifying the generated data as belonging to the original dataset and ultimately minimize V(D, G) which is the cost value function.

During training, the generator encapsulates the probability distribution of original data and is trained to generate data that maximizes the probability of the discriminator mistaking the fake data to be real. The end goal of the generator is such that the discriminator is no longer able to differentiate between real data or synthetically generated data. The generator module is a neural network that consists of one input layer, one or more hidden layers, and an output layer. The discriminator module in GAN can be thought of as a classifier. The aim of the discriminator is to categorize data under analysis to real or fake. The architecture of the discriminator depends on the type of data it is classifying.

The combined loss of the GAN can be represented by the following equation -

$$\min_G \max_D V(D, G) = \min_G \max_D \left( E_{x \sim P_{data}(x)}[log D(x)] + E_{z \sim P_z(z)}[log(1 - D(G(z)))] \right)$$

COVID-CT-Dataset[1] identified the difficulty of publically available COVID-19 CT datasets due to various privacy issues and came up with its own publicly available dataset. The dataset is composed of 349 chest CT scans of COVID-19 positive patients collected from 216 patients and it contains 463 non-COVID-19 CT images. It has also verified this dataset from a senior radiologist. It also came up with useful experimentation results where it demonstrated the usefulness of the dataset by building AI models for diagnosing COVID-19. The diagnosis model achieves impressive results with an accuracy of 0.89 by exploiting multi-task learning and self-supervised learning.

## III. METHODOLOGY

During the analysis of the dataset, we realized that the images available in the dataset are of different dimensions and have different brightness. So, the first step in our experiment was to preprocess the images and transform them to a fixed size and normalize the brightness. We decided to resize the images to 224 x 224 for our models. Using the PyTorch Vision library, we performed the following on the images—downscaling, random resized crop, random horizontal flip, and normalization.

We chose Deep Convolutional GANs (DCGANs) for generating images as our generative model. We designed the Discriminator and the Generator model such that their architectures are symmetric since it is the simplest and the most effective way of ensuring that both models are equally powerful and have fair competition. The complete architecture is trained using the PyTorch framework.

The generator takes input as a noise vector of shape 100 x 1 and outputs a single 224 x 224 x 3 image. The first layer of the network is a fully connected layer with 100 input features and 150528 output features. The output of this layer is provided as input to five transpose convolutional layers (ConvTranspose2D) to upsample the input vector, first to 3072 x 7 x 7, then 1536 x 14 x 14, then 768 x 28 x 28, then 384 x 56 x 56, then 192 x 112 x 112, and finally to 3 x 224 x 224. Except for the last layer, each convolution transpose layer is followed by a batch normalization layer (BatchNorm2D) and a Rectified unit (ReLU) activation layer. The transpose convolutional layers are configured to use a kernel size of (4, 4) and stride of (2, 2). The activation layer uses the ReLU function whereas the output layer uses the hyperbolic tangent (Tanh) function. The generator has ~30 million parameters. The output of the generator is an image of shape 224 x 224 x 3. The layer-based architecture of the Generator and Discriminator model is shown in Fig. 1 and Fig 2.

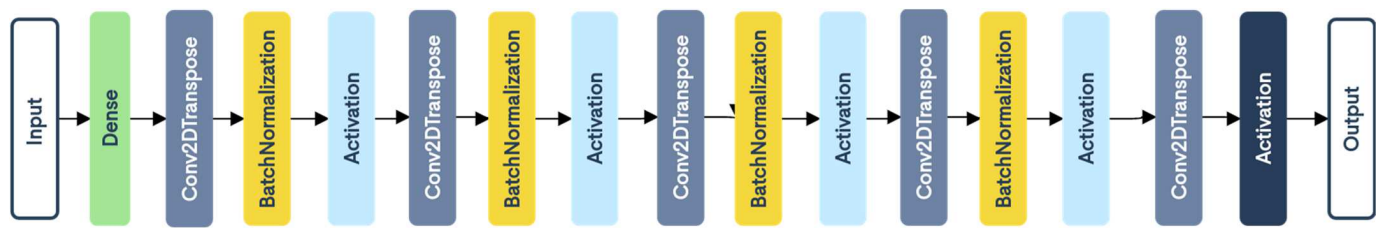

**Fig 1:** Layered architecture of the Generator model

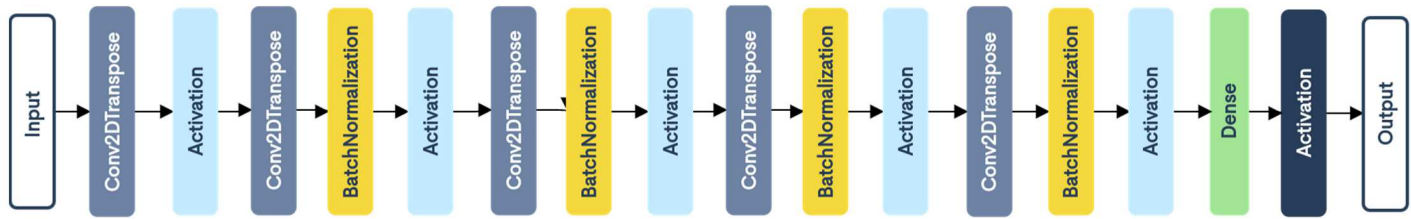

**Fig 2:** Layered architecture of the Discriminator model

A CNN based model is used as the Discriminator model which outputs whether the image is fake/real (class = 0/1). The input image is passed through five convolutional layers (Conv2D) each of which is followed by a batch normalization layer (BatchNorm2D and a LeakyReLU activation layer. The input is downsampled from 224 x 224 x 3 to 192 x 112 x 112, then 384 x 56 x 56, then 768 x 28 x 28, then 1536 x 14 x 14, then 3027 x 7 x 7. Kernel size chosen for the model is (4, 4), the size of the stride chosen is (2, 2) and the activation function used is LeakyReLU with a slope of 0.2. There are approximately 5 million parameters in the Discriminator. The final output is reduced to a vector and passed to a dense layer. The output of the dense layer is used to estimate if the image is real or not. The output layer predicts the realness using the Sigmoid function. The layered architecture of the Generator and Discriminator models are shown in Fig. 3 & 4 respectively.

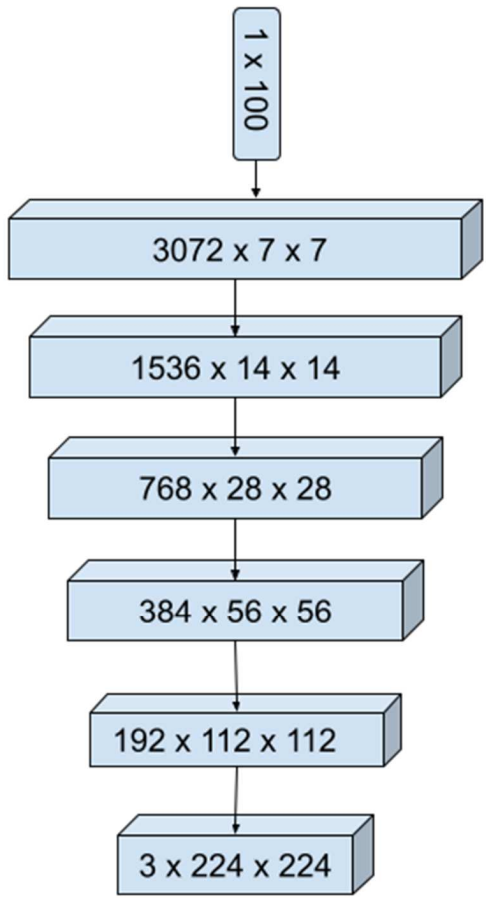
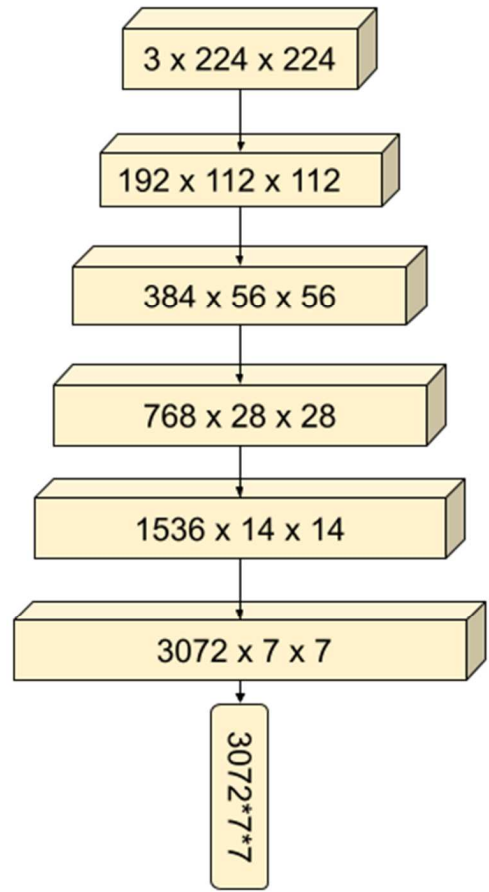

**Fig 3:** Generator architecture      **Fig 4:** Discriminator architecture

The loss function used for the GAN is the "binary cross-entropy loss". The loss can be represented by the simple equation -

$$L = \min_{G} \max_{D} [log(D(x)) + log(1 - D(G(z)))]$$

The major issue of GANs is the validation of the generated image i.e. the generated image is accurate or not. This task might be trivial in cases like Face Generation or Image to Image translation but not in this case. Generated images need to be validated on whether they are actually of COVID-19 positive Chest CT Scans. The ideal way would be to get a radiologist to review the generated images and handpick the images that can be added to our dataset. But since this is not possible, we decided to use a Baseline model[3] that can perform the same task. We used a CNN based model available with the dataset to measure the performance of the GAN model. The baseline model is provided by the COVID-CT repository[1]. The baseline model is an optimized version of the DenseNet-169 pre-trained model and it classifies the input images in two classes - COVID-19 positive and Normal.

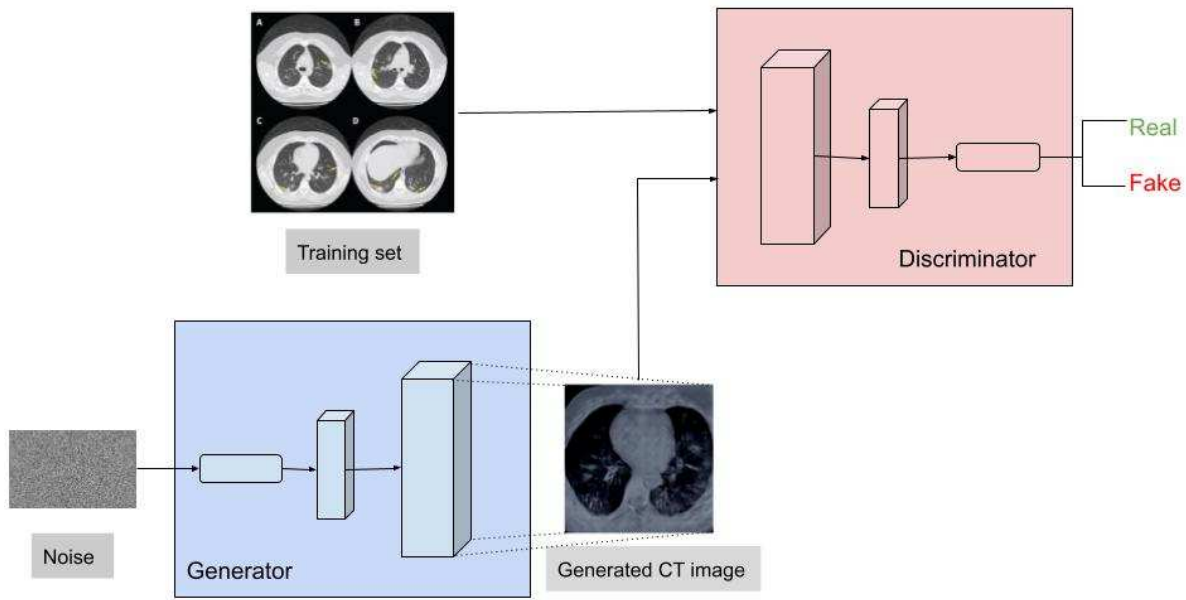

**Fig 5:** Model Training

IV. EXPERIMENTS AND RESULTS

The model is trained to generate COVID-19 positive Chest CT images. The image preprocessing involves resizing the image to 224 x 224 x 3 and then normalizing the image pixels to [-1, 1] from [0, 255]. Adam is used as the optimizer function. The hyperparameters used for the training are as follows: "batch size: 16, learning rate: 0.0002, beta: 0.5 (Adam optimizer momentum) and the number of epochs: 3000." The model has ~35 million parameters.

The images generated by our DCGAN were compared with the Baseline model provided by the COVID-CT repository[1]. The baseline model is an optimized version of the DenseNet-169 pre-trained model. We generated ten sets of 100 COVID-19 positive chest CT images using the Generator model and predicted their nature using the baseline model. Since all images should be COVID-19 positive the accuracy is calculated by counting the number of images marked COVID-19 positive by the baseline model. On average, the model predicted about 40% of generated images to be COVID-19 positive. The generated images and calculated accuracies for the sets can be seen in Fig. 6 & 7 respectively.

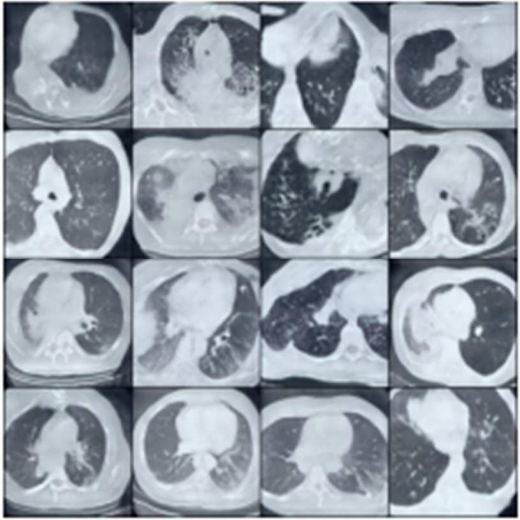 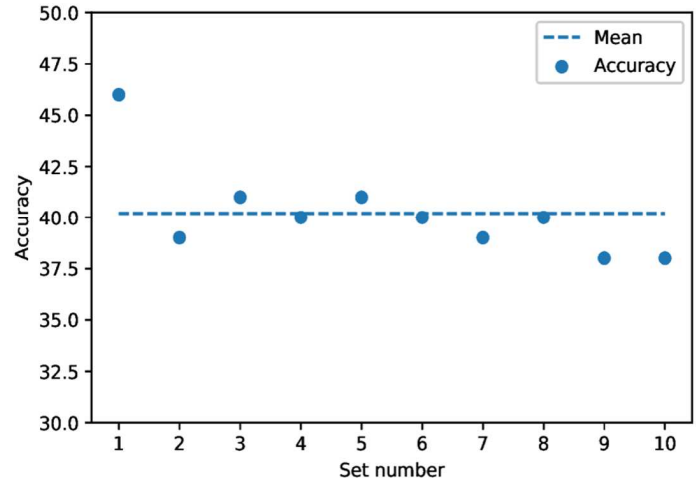

**Fig 6**: Generated Chest CT images      **Fig 7**:Scatter graph of accuracy vs Set.

## V. CONCLUSION AND FUTURE WORK

In this paper, we have shown how GANs can be useful to generate synthetic images. Using the baseline models we can see that around 40% of images are being correctly predicted as being COVID-19 positive. The motive behind generating synthetic images was to extend the existing dataset of chest CT images which can be utilized to build a CNN-based predictive model. The models that are currently available for detection of SARS-CoV-2 using chest CT scans [5][6][7] were trained on small datasets of chest CTs and have accuracy in the range of 85-90%. CNN-based networks are prone to overfitting if trained using a small dataset. Therefore, if these available networks are trained on the extended dataset, they could perform better and give more accurate results. Also, the extended dataset could be published separately for the development of further predictive models.